\documentclass[runningheads]{llncs}
\usepackage{graphicx}
\usepackage{multirow}
\usepackage{lscape}
\usepackage{rotating}
\usepackage{xcolor}

\begin{document}
\title{Neural Hybrid Recommender: Recommendation needs collaboration
}

\author{Ezgi Y{\i}ld{\i}r{\i}m\inst{1,2} \and
Payam Azad\inst{2} \and
\c{S}ule G\"{u}nd\"{u}z \"{O}\u{g}\"{u}d\"{u}c\"{u}\inst{1}}
\authorrunning{E. Y{\i}ld{\i}r{\i}m et al.}
\institute{Istanbul Technical University, 34467 Sar{\i}yer/Istanbul, Turkey \\
\email{\{yildirimez,sgunduz\}@itu.edu.tr}\and
Catharijnesingel 30e, 3511 GB Utrecht, Netherlands
\email{\{ezgi.yildirim,payam.azad\}@5ca.com}}
\maketitle
\begin{abstract}
In recent years, deep learning has gained an indisputable success in computer vision, speech recognition, and natural language processing. After its rising success on these challenging areas, it has been studied on recommender systems as well, but mostly to include content features into traditional methods. 
In this paper, we introduce a generalized neural network-based recommender framework that is easily extendable by additional networks. This framework named NHR, short for \textit{Neural Hybrid Recommender} 
allows us to include more elaborate information from the same and different data sources. We have worked on item prediction problems, but the framework can be used for rating prediction problems as well with a single change on the loss function. To evaluate the effect of such a framework, we have tested our approach on benchmark and not yet experimented datasets. The results in these real-world datasets show the superior performance of our approach in comparison with the state-of-the-art methods.

\end{abstract}

\section{Introduction}
\label{sec:intro}
Online services such as social media and e-commerce have played the key role to derive massive data sources for information systems. Since this information explosion makes users' lives more complicated and even difficult to use such systems, recommender systems aim to offer personalized recommendations to users in order to minimize confusion and increase the chance to reach meaningful information. 
Based on the available data and the nature of the application domain, there are two main approaches in recommender systems to produce favorable recommendations: collaborative filtering that learn only from past interactions of users and content-based methods that learn the taste of users by using content features. 
However, both approaches have flaws and favors. While collaborative filtering does not require domain expertise to mine information from data sources and works well for complex objects such as movies, books, music, etc. where variations in taste are much sparse than variations in preferences; content-based filtering works better if preference data is sparse and cold-start is an issue. In practice, companies are following a middle way and using hybrid systems of these two approaches. Nevertheless, there are seldom cases of hybrid recommender systems investigated in the literature. Therefore, we present a general framework to use both aspects in a compact deep neural network architecture.

Among the various applied methods, matrix factorization is the most known collaborative filtering approach. Matrix factorization projects user and item into a shared latent space by decomposing the rating matrix into low-dimensional latent factors. To find out an interaction between user and item, the inner product of latent factors are used in recommender systems. In \cite{li2015deep}, a deep collaborative filtering (DCF) method is proposed to combine probabilistic matrix factorization (PMF) with marginalized denoising auto-encoders (mDA). The latent factors are extracted from the hidden layer of deep networks and they are used to feed matrix factorization components. A collaborative topic modeling approach is proposed by Wang and Blei \cite{wang2011collaborative} for recommending scientific articles to online communities. Here, Latent Dirichlet Allocation (LDA) is applied to the user ratings as well as the article contents. Once users and articles are represented as latent factors, matrix factorization is applied to their latent representations to predict user preferences. \cite{kim2016convolutional} proposed a context-aware recommendation model, convolutional matrix factorization (ConvMF) that integrates a convolutional neural network (CNN) into PMF. Item representation is obtained from the CNN network that they have trained directly in matrix factorization. 

In most of the studies in recommender systems, Deep Neural Networks (DNNs) are used to either get better latent factor representation or integrate auxiliary information into matrix factorization to alleviate the cold-start problem. In contrast to the wide range of researches on the combination of matrix factorization and DNNs, there is relatively little work on employing DNNs to learn the interaction function directly from data. A very first attempt to build a traditional collaborative filtering setup by neural networks \cite{dziugaite2015neural} simulated matrix factorization by replacing its inner product by a feed-forward neural network, however, it could not be succeeded in benchmark datasets. \cite{he2017neural} took this approach one step further because the inner product cannot capture non-linear interactions between users and items. Thus, they proposed a framework named NCF to replace the inner product with non-linear interaction function by a feed-forward neural network and they reported promising results. However, interaction data by itself cannot be sufficient for a challenging recommender system in most cases, auxiliary data is a key factor especially for the systems introducing new users or items at any time. 
This paper explores the use of DNNs to extract meaningful information from both auxiliary and historical interaction data, then combines them to make better predictions than any single aspects and data sources. Our proposed framework can be extended by not yet experimented auxiliary data and/or by redefining the interaction function using the current data in a flexible manner.

The main contributions of this work are summarized below.
\begin{itemize}
    \item We devise a general framework for a hybrid recommender system based on DNNs that model latent features of user and item from both auxiliary and interaction data.
    \item We demonstrate the effectiveness of our NHR approach on the collaboration of self-sufficient recommender models. 
    \item We verify that auxiliary information can significantly improve recommendation quality, especially in large-scale domains. Utilizing auxiliary information can improve not only the success in detecting true interactions but also the ability to correctly rank predictions.
    \item 
    We show that our NHR approach is essential in the domains that suffer from the severity of cold-starts and rating sparsity due to its stronger contributions to such disadvantaged domains.
\end{itemize}

Recommendation problems generally suffer from the lack of actual feedbacks given by users a.k.a. explicit feedback. Explicit feedback (via ratings and reviews) is a clear expression of user preferences on items, and it is expressed by direct interactions between system and user. On the other hand, implicit feedback is automatically tracked by the system itself, through inferences about the behavior of the user, such as watching videos, purchasing products and clicking items. Despite the plethora of research over explicit feedbacks; implicit feedbacks are the more realistic case of recommender systems in uttermost situations such as online advertising and online shopping. The reason for the less popularity of using implicit feedbacks is its challenging nature due to the absence of negative interactions. Since we have tested our framework on item prediction problems, we employ negative sampling as discussed in Section \ref{sec:negative} to come through this problem. 

   
\section{Neural Hybrid Recommender} \label{sec:nhr}
\begin{figure}[t]
	\centering
	\includegraphics[width=\linewidth]{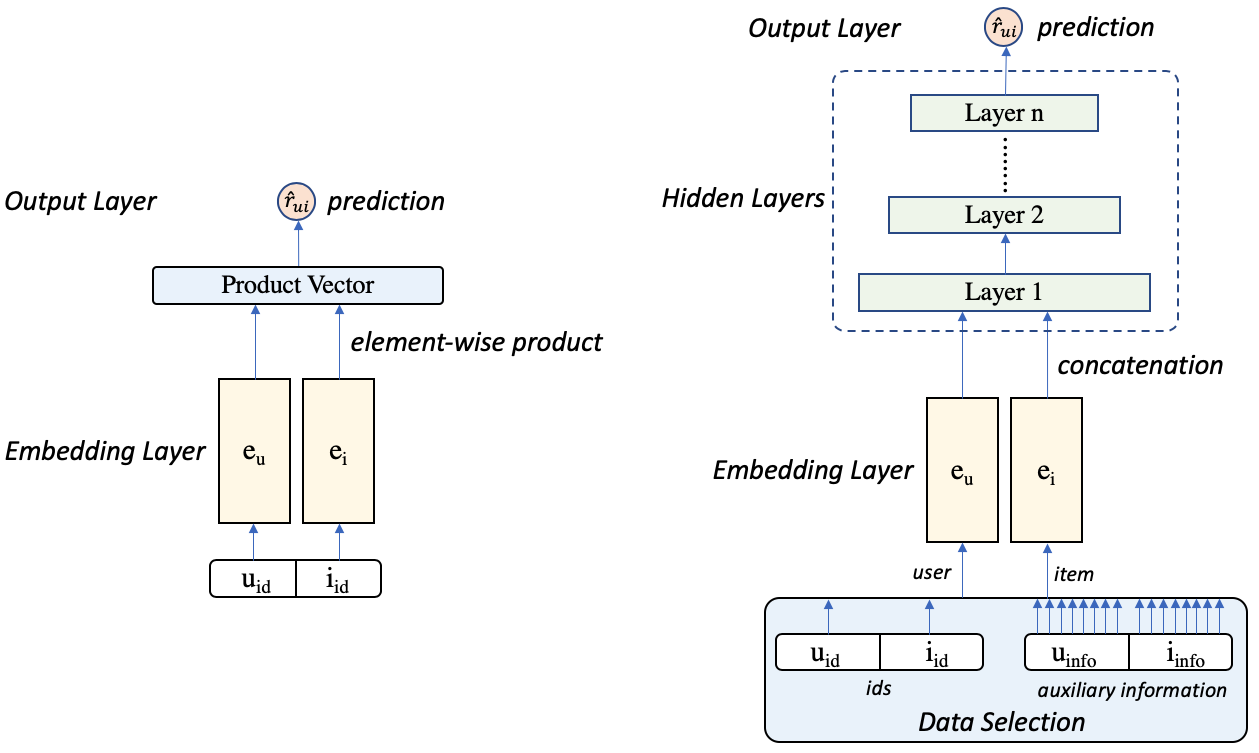}
	\caption{(left) Representation of neural network realization of matrix factorization; (right) Representation of deep neural recommender networks}
	\label{fig:mf_and_mlp}
\end{figure}

In order to build a general framework for both collaborative filtering and auxiliary information, we adopt feed-forward neural networks. Neural networks can model user-item interaction since it has been proven that they are able to learn non-linear relations which is essential for the recommendation of complex objects such as jobs and movies. As suggested in \cite{cheng2016wide}, we also utilize wide neural networks for memorization of feature interactions through a wide set of cross-product feature transformations and deep neural networks for better generalization of unseen feature combinations through low-dimensional dense embeddings. 
Following NCF, we first build a Wide\&Deep collaborative filtering approach by combining different neural networks using the same interaction data, then we add auxiliary information by supplementary networks into the system to address the cold-start problem. The names of pure collaborative filtering methods remained as in \cite{he2017neural}: GMF (Generalized Matrix Factorization) performing non-linear matrix factorization and MLP (Multi-Layer Perceptron) learning the high-order interaction function. The models trained on auxiliary information are simply named NHR-\textit{type} where \textit{type} refers to the data type that is used for training.
We first train multiple self-sufficient neural recommenders independent from each other, then build a framework as an ensemble of all 
. Even though there is no limitation on the construction of the models, we can roughly divide what type of networks we use in our experiments into two groups:
\begin{itemize}
    \item neural network realization of matrix factorization 
    (Fig. \ref{fig:mf_and_mlp}-left)
    \item deep neural recommender networks 
    (Fig. \ref{fig:mf_and_mlp}-right)
\end{itemize}

%

\begin{figure}[t]
	\centering
	\includegraphics[width=0.95\linewidth]{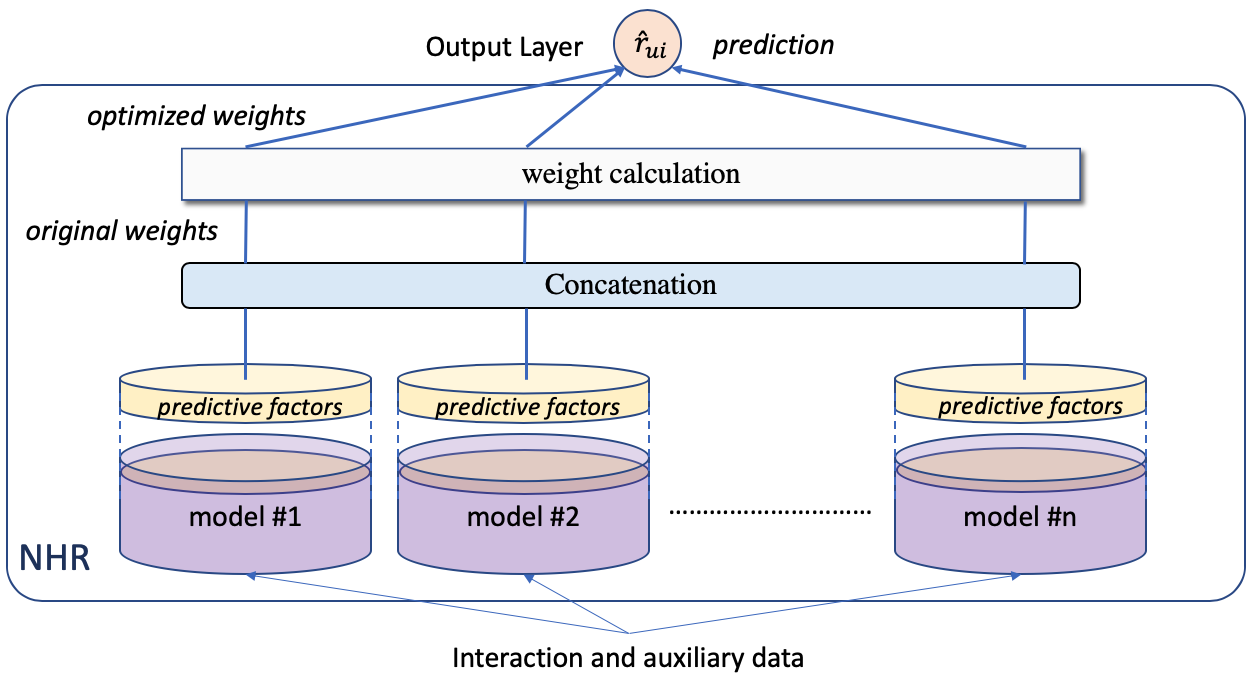}
	\caption{Network Architecture for Neural Hybrid Recommender Framework 
	}
	\label{fig:nhr}
\end{figure}

Both of the mentioned networks have embedding layers to transform users and items into vector representations. The obtained embedding vectors can be interpreted as the latent vectors of users and items. If we term $p_{u}$ and $q_{i}$ as the user latent vector and item latent vector respectively, one can easily define a mapping function as
\begin{equation}
    \phi_{mul}\left(p_{u}, q_{i}\right)=p_{u} \odot q_{i}
\end{equation}
where $\phi$ denotes the element-wise product of latent vectors. Then, the next step is to project this product vector to the output layer of the model:
\begin{equation}
    \phi_{out}\left(x\right)=\alpha_{out}\left(W^{T}x+b\right)
\end{equation}
where $x=\phi_{mul}\left(p_{u}, q_{i}\right)$, the output of the multiplication layer in Fig. \ref{fig:mf_and_mlp}-left, and $W$, $b$ and $\alpha_{out}$ is the weight vector, bias, and activation function of the layer, respectively. Under the assumptions that the weight vector $W$ is a uniform vector of 1, there is zero bias $b$ in the equation and the activation is an identity function which allows firing the perceptron with the exact value of the input, this project layer acts as a traditional matrix factorization. In order to implement neural network realization of matrix factorization, the weight vector $W$ and the bias $b$ are learned from interactions by the logarithmic loss function in Eq. \ref{eq:binary-cross-entropy}, and in this way, a non-linear MF approach a.k.a. GMF is obtained. The sigmoid function $\sigma(x)=1/\left(1+e^{-x}\right)$ is used as $\alpha_{out}$ because it restricts each neuron to be in $\left(0,1\right)$ range which meets the expectation for item prediction.


The outputs of the embedding layers on GMF and MLP models are already 1-dimensional vectors because they are fed on inputs of length 1 (\textit{id}s only). However, the embedding layers of deep neural recommender networks trained on auxiliary data (NHR) produce sequences of embeddings w.r.t. sequence length. Average-pooling is a well-known application to gather information exists in the sequence members into a particular form, for example getting sentence embeddings from word embeddings \cite{wieting2015towards,adi2016fine}, average-pooling is applied to the outputs of embedding layers in these models. Since users and items are represented with several features and every feature has its own embedding space, a concatenation is applied to have one unique latent vector representation for each user-item pair after the average-pooling of embeddings. 

Once the latent vectors are obtained for user-item pairs, the following functions are used to generate MLP and NHR models.
\begin{equation}
    \begin{array}{c}
    {\phi_{1}\left(\phi_{concat}\right)=\alpha_{1}\left(W_{1}^{T}\phi_{concat}+b_{1}\right)} \\
    {\phi_{2}\left(\phi_{1}\right)=\alpha_{2}\left(W_{2}^{T}\phi_{1}+b_{2}\right)} \\
    \vdots \\
    {\phi_{n}\left(\phi_{n-1}\right)=\alpha_{n}\left(W_{n}^{T}\phi_{n-1}+b_{n}\right)} \\ \end{array}
\end{equation}
where $\alpha_{x}$s are ReLU activation functions, except the final $\alpha_{n}$ which is a sigmoid. $W_{x}$s are the weight matrices and $b_{x}$s are bias vectors as usual.

As reported in \cite{erhan2010does}, the initialization of weights can contribute to convergence and performance of deep learning models. Therefore, we first train all models without prior information till the convergence, then use their parameters to initialize relevant weights on the overall architecture. To combine the models, we simply concatenate the last layers of networks just before the outputs. Since this layer defines the predictive capability of a model, it is generally called as \textit{predictive factors} in literature. We use the original weights of last layers in a weighting process:
\begin{equation}
w \leftarrow \left[ \begin{array}{c}
{\alpha w^{1}}\hspace{0.2cm}
{\beta w^{2}}\hspace{0.2cm}
{...}\hspace{0.2cm}
{\gamma w^{n}}
\end{array}\right] \quad where\  (\alpha+\beta+...+\gamma)=1
\end{equation}
where $w^{n}$ denotes the weight vector of $n$th pre-trained model and ($\alpha$, $\beta$, ..., $\gamma$) is the set of hyper-parameters determining the trade-off between the pre-trained models. The final framework which ensembles multiple self-sufficient neural recommender networks by this weighting process is shown in Fig. \ref{fig:nhr}. 

The parameters given in the layer definitions of all models are learned by binary cross entropy loss function given below.
\begin{equation}
\label{eq:binary-cross-entropy}
\mathcal{L}=-\sum_{(u, i) \in \mathcal{O} \cup \mathcal{O}^{-}} r_{u i} \log \hat{r}_{u i}+\left(1-r_{u i}\right) \log \left(1-\hat{r}_{u i}\right)
\end{equation}
where $\mathcal{O}$ denotes the set of observed interactions, and $\mathcal{O}^{-}$ denotes the set of negative instances. When the loss function is replaced to a weighted squared loss, the proposed framework can be easily applied to explicit datasets as well.

\section{Experiments}
\label{sec:experiments}
\subsection{Datasets}
\label{sec:datasets}
To conduct our experiments, we worked on two real-world problems: movie recommendation and job recommendation. For the movie recommendation task, we applied our approach to a benchmark movie rating dataset enriched by movie subtitles.

\subsubsection{MovieLens \& OPUS.}
\label{sec:datamovie}
MovieLens 
\cite{harper2016movielens} includes 5-star ratings of movies and some categorical properties of users and movies. It contains $1M$ ratings, $3.8K$ movies and $6K$ users in total. Users have at least 20 ratings. 5-star explicit ratings are converted to implicit feedback by treating a rating is the indicator of user-item interaction, so all ratings in the dataset are considered to be 1. OPUS subtitles dataset 
\cite{lison2016opensubtitles2016} describes a collection of translated movie subtitles from \url{http://www.opensubtitles.org/}. It composes of bitexts from many language pairs.
English subtitles are used to supply more convenient contents for movies. 2581 movies out of 3706 (69.64\%) in the rating dataset have subtitles.
The movie subtitles in the OPUS dataset are utilized for item representation while and the categorical properties of user profiles in the MovieLens for user representation.

\begin{table}[t]
\centering
\caption{Statistics of the experimented datasets}
\label{tbl:datasets}
\begin{tabular}{cccccc}
\textbf{Dataset} & \textbf{Type} & \textbf{Interaction} & \textbf{Item} & \textbf{User} & \textbf{Sparsity} \\ \hline
MovieLens & movie & 1,000,209 & 3,706 & 6,040 & 95.53\% \\
Kariyer & job & 383,434 & 16,134 & 20,283 & 99.88\%
\end{tabular}
\end{table}

\subsubsection{Kariyer.}
\label{sec:datajob}
This dataset consists of the job application history of candidates from a one-week period, candidate profiles, job definitions, job requirements, company details. Each user has at least 20 applications. It consists of $383K$ applications, $20K$ candidates for $16K$ jobs in total. The application history of users is used as the interaction data in job recommendation, and the properties of jobs and candidates as the auxiliary data.

\subsection{Handling Text Data}
\label{sec:text}
To make the text data suitable to feed neural networks, we need to convert raw texts into numeric vectors. 
In the simplest approach, using a simple dictionary for this purpose could lead to extremely sparse representations due to the huge size of vocabulary. Thus, we exploited the advantage of a hash function which converts a raw text to a sequence of indexes in a fixed-size hashing space. Note that some words may be assigned to the same index according to the hash function. The dimension of hashing space is in relation to the overlapping rate of distinct words and the dimension of embedding layers. 
By considering the pros and cons, we set the dimension of hashing space to $1K$ in the experiments after evaluating its effect on overall performance and complexity. 

Since the inputs to the neural networks have to be in the same size for all iterations, 
we examined the mean ($\mu$) and the standard deviation ($\sigma$) of sequence lengths of text features. Then, the feature-specific input lengths are defined as $\mu + \sigma$ for each text feature in the datasets.

\subsection{Evaluation Process}
\label{sec:evaluation}
In order to split the dataset into the train and test sets, we preferred \textit{leave-one-out} evaluation which has been widely applied in many works \cite{he2016fast,he2017neural,bayer2017generic,rendle2009bpr}, especially where sparse datasets are subjected. The latest interaction of each user is held-out to compose a test set, while the remaining interactions are used for training. The last interaction of each user in the train set is used for hyper-parameters tuning.

Since ranking every user-item pair amongst the test pairs are very time-consuming and not possible to run in real-time. Therefore, as in similar studies \cite{koren2008factorization,elkahky2015multi,he2017neural} we randomly sampled 100 items per user and rank them by probability of interaction. To measure the quality of ranking, we used well-known evaluation metrics: Hit Ratio (HR) and Normalized Discounted Cumulative Gain (NDCG). We applied both metrics on a truncated list including top-10 ranked test items for each user. Due to the fact that the users have one interaction in the test set, HR@k is simplified in our experiments as follows:
\begin{equation}
\label{eq:hr}
HR@k=\left\{\begin{array}{ll}{1/k,} & {\mathrm{if} \hspace{0.05 in} r_{test}(u,i) \in R_{k}} \\
{0,} & {\mathrm {otherwise}}\end{array}\right.
\end{equation}
where $r_{test}(u,i)$ and $R_{k}$ define the interaction with the item $i$ and the list of top-$k$ recommended items for the user $u$. In addition to HR@k, NDCG@k is reinterpreted as well in our experiments because ideal discounted cumulative gain ($IDCG_{k}$) in position $k$ is equal to $1$ in our evaluation setup. Therefore, NDCG@k is redefined as:

\begin{equation}
\label{eq:ndcg}
NDCG@k=\frac{DCG@k}{IDCG@k}=\sum_{i=1}^{k} \frac{2^{r(u,i)}-1}{\log(i+1)}
\end{equation}
where $r(u,i)$ is $1$ if the user $u$ interacted with the $i$th item of the top-$k$ list and $0$ otherwise. The results are reported by the mean of user scores.

HR gives a shallow understanding of success by considering if the interacted item is in the top-10 list or not whereas NDCG helps for a better understanding by setting higher scores to hits at higher ranks. 

\subsection{Negative Sampling}
\label{sec:negative}
In most of the cases, implicit feedback refers to positive inference of user interaction or user interest.
To handle the absence of negative feedback, many studies have either assumed all unobserved cases as negative feedback or sampled negative instances from them. In this work, we also apply the latter approach to generate a set of negative feedback by sampling four negative instances per positive instance. 
Unlike the evaluation process, we randomly sampled negative training instances in real-time, just before each epoch starts. This allows our system to learn as much as possible from different instances and increases the utility of dataset without interfering with its feasibility.  



\subsection{Baselines}
\label{sec:baselines}
We compared our proposed approach NHR to the following methods:
\begin{itemize}
    \item \textbf{PopRank} is a non-personalized popularity based recommendation method. Items are ranked by their popularity which is determined by the number of interactions and recommended to all users with the same order.
    \item \textbf{BPR} \cite{rendle2009bpr} is a highly competitive pairwise ranking method which works well for implicit feedbacks. It optimizes the matrix factorization model with a pairwise ranking loss.
    \item \textbf{ALS} \cite{hu2008collaborative} is also a matrix factorization algorithm for item recommendation. It works in parallel and effective for large-scale collaborative filtering problems which suffer from the sparseness of the rating data. 
    \item \textbf{GMF} \cite{he2017neural} is a neural network realization of matrix factorization. Besides being a part of NCF, it can be employed as a complete recommender system.
    \item \textbf{MLP} \cite{he2017neural} is also a part of NCF that learns user-item interaction function by neural networks, Like GMF, it is a standalone recommender system.
    \item \textbf{NCF} \cite{he2017neural} is a state-of-the-art neural network based collaborative filtering method which combines GMF and MLP methods. No matter that has very promising results for item prediction, it is a pure collaborative filtering method which benefits from only interaction data and does not regard cold-starts that is a very common case for real-world recommendation tasks.
\end{itemize}


\subsection{Parameter Setting}
We implemented our proposed framework using PyTorch. 
All individual models had been learned by optimizing the logarithmic loss of Eq. \ref{eq:binary-cross-entropy} because we tested them on an item prediction setup. 
To determine the hyper-parameters of the methods, we conducted intensive tests on validation data.

For individual models that are trained without any prior information, we set model parameters with a Xavier initialization
, then optimize them with Adam optimizer 
which employs an adaptive learning rate for faster convergence. The learning rate is set to 0.001 and the momentum for Adam optimizer to 0.9 which is the default setting.

We tested a bunch of different batch size but found the 128 is the best performing setup for all, except the model trained on text data. Because the embedding size for the text data is quite large and hard to fit on even comparatively large computer memories, we adopt the batch size of 32 for them. 

We evaluated the predictive factors of $\{8, 16, 32, 64\}$. We employed three hidden layers for interaction-specific networks, for example, if the number of predictive factors is set to $8$, then the size of hidden layers are selected in the order of $32 \rightarrow 16 \rightarrow 8$ from the top on down and the embedding size is $16$ in this setup, as a matter of course. For the networks trained on auxiliary data, we used two hidden layers and intuitively set embedding size to be $128$ for movie subtitles, $4$ for job titles and candidate past-positions, and $16$ for job qualifications, job explanations and candidate experiments. 
To treat equally, we set the $\alpha$ parameter of NCF which defines the trade-off between GMF and MLP by optimization as we did for our NHR methods. 

\begin{table*}[t]
\centering
\caption{Performance of HR@10 and NDCG@10 w.r.t. the number of predictive factors (\textit{pf}) on different datasets. Here are the abbreviations used to shrink the result table due to the limited space; \textit{ds}:Dataset, \textit{ML}:MovieLens, \textit{Ka}:Kariyer, \textit{mt}:Metric, \textit{PR}:PopRank, \textit{cat.}:categorical, \textit{comb.}:combined, and \textit{Im.}:Improvements}
\label{tbl:results}
\begin{tabular}{c|c|c|c|c|c|c|c|c|c|c|c|c} 
\hline\hline
\multirow{2}{*}{\textbf{ds}} & \multirow{2}{*}{\textbf{pf}} & \multirow{2}{*}{\textbf{mt}} & \multicolumn{6}{c|}{\textbf{Baselines} }                                                     & \multicolumn{3}{c|}{\textbf{NHR}}                   & \multirow{2}{*}{\textbf{Im.\%}}  \\ 
\cline{4-12}
                              &                               &                               & \textbf{PR}  & \textbf{BPR}  & \textbf{ALS}  & \textbf{GMF}  & \textbf{MLP}  & \textbf{NCF}  & \textbf{cat.}    & \textbf{text}  & \textbf{comb.}   &                                   \\ 
\hline
\multirow{8}{*}{ML} & \multirow{2}{*}{8} & HR & 0.4512 & 0.5331 & 0.6076 & 0.6247 & 0.6522 & 0.6560 & - & - & \textbf{0.6718} & 2.41\% \\
 & & NDCG & 0.2546 & 0.3027 & 0.3488 & 0.3528 & 0.3789 & 0.3807 & -      & -      & \textbf{0.3943} & 3.57\% \\ \cline{2-13}
 & \multirow{2}{*}{16} & HR & 0.4512 & 0.5886 & 0.6545 & 0.6714 & 0.6626 & 0.6828 & -      & -      & \textbf{0.6946} & 1.73\% \\
 & & NDCG & 0.2546 & 0.3426 & 0.3886 & 0.3945 & 0.3890 & 0.4057 & -      & -      & \textbf{0.4126} & 1.7\% \\ \cline{2-13}
 & \multirow{2}{*}{32}& HR  & 0.4512 & 0.6040 & 0.6826 & 0.6757 & 0.6728 & 0.6874 & -      & -      & \textbf{0.6979} & 1.53\% \\
 & & NDCG & 0.2546 & 0.3564 & 0.4150 & 0.3936 & 0.3986 & 0.4053 & -      & -      & \textbf{0.4147} & 2.32\% \\ \cline{2-13}
 & \multirow{2}{*}{64}& HR  & 0.4512 & 0.6108 & 0.6912 & 0.6763 & 0.5190 & 0.6798 & -      & -      & \textbf{0.6964} & 2.44\% \\
 & & NDCG & 0.2546 & 0.3621 & 0.4290 & 0.4052 & 0.2857 & 0.4077 & -      & -      & \textbf{0.4176} & 2.43\% \\ 
\hline\hline
\multirow{8}{*}{Ka} & \multirow{2}{*}{8} & HR & 0.3231 & 0.7399 & 0.5137 & 0.8249 & 0.7448 & 0.8594 & 0.8821 & 0.8624 & \textbf{0.8834} & 2.79\% \\ 
 & & NDCG & 0.1875 & 0.5067 & 0.3237 & 0.5719 & 0.5592 & 0.6204 & \textbf{0.6368} & 0.6188 & 0.6354 & 2.64\% \\ \cline{2-13} 
 & \multirow{2}{*}{16} & HR & 0.3231 & 0.7874 & 0.6166 & 0.8357 & 0.8021 & 0.8695 & 0.8890 & 0.8730 & \textbf{0.8917} & 2.55\% \\ 
 & & NDCG & 0.1875 & 0.5560 & 0.4034 & 0.6041 & 0.5564 & 0.6402 & 0.6571 & 0.6426 & \textbf{0.6579} & 2.76\% \\ \cline{2-13} 
 & \multirow{2}{*}{32} & HR & 0.3231 & 0.7934 & 0.7013 & 0.8121 & 0.8100 & 0.8658 & 0.8851 & 0.8703 & \textbf{0.8875} & 2.51\% \\ 
 & & NDCG & 0.1875 & 0.5629 & 0.4740 & 0.5870 & 0.5471 & 0.6369 & 0.6537 & 0.6411 & \textbf{0.6562} & 3.03\% \\ \cline{2-13} 
 & \multirow{2}{*}{64} & HR & 0.3231 & 0.7922 & 0.7627 & 0.7841 & 0.8205 & 0.8621 & 0.8800 & 0.8678 & \textbf{0.8841} & 2.55\% \\ 
 & & NDCG & 0.1875 & 0.5608 & 0.5394 & 0.5624 & 0.5519 & 0.6334 & 0.6505 & 0.6378 & \textbf{0.6536} & 3.19\% \\ \hline
\hline
\end{tabular}
\end{table*}

\subsection{Performance Results}
\label{sec:results}
In our NHR experiments, we group auxiliary information sources into three categories: categorical, text, and a combination of them. Kariyer dataset includes many data types: free-text, real values, binary, single-label, and multi-label categorical features. 
In order to handle all different types during the learning process, we first apply general pre-processing steps such as outlier removal, tokenization, etc. We then normalize real values and transform binary and categorical features into one-hot and multi-hot representations. All these features are considered \textit{categorical} for simplicity. We also convert raw text features to hash vectors which refer to \textit{text} data source as explained in Section \ref{sec:text}; 
Both networks trained on the categorical and text data sources are first incorporated into NCF alone (NHR-categorical and NHR-text respectively), then together to embody the most extensive NHR model (NHR-combined). 
As for MovieLens dataset, users are represented with categorical features whereas movies are represented with text features. 
This results in having one auxiliary network (NHR-combined) which combine the categorical and the text data sources at the same time. Thus, we could report one experiment on NHR for the movie recommendation task.

Table \ref{tbl:results} shows the recommendation performance of the compared methods with respect to the number of predictive factors. The results are given in HR@10 and NDCG@10. BPR and ALS methods have the same latent factor size as the predictive factors in neural network models. By doing so, we use the same predictive capability for all baselines except PopRank to make a fair comparison between them. PopRank has the weakest performance amongst the other methods. It is already expected because it is incapable to make personalized suggestions. 
Since 0.001-level improvements are already found to be significant for similar tasks such as click-through rate (CTR) prediction \cite{cheng2016wide,wang2017deep,guo2017deepfm,song2018autoint}, one can easily say that NHR is significantly outperforming the state-of-the-art matrix factorization methods, ALS and BPR, by a large margin in both metrics, and it is also consistently superior to the most competitive baseline NCF. NHR on MovieLens and Kariyer achieved 2.03\% HR-2.51\% NDCG and 2.60\% HR-2.91\% NDCG relative improvements on average over their NCF counterparts, relatively. NHR gains more generalization capability through merging 
interaction and auxiliary data. In addition to more accurate hits on top-10 predictions, the results show that NHR systems could better learn to rank items in the top-10 lists by uprising the test interaction amongst the other predictions since NDCG scores are improved by larger steps.
The NHR-combined results on job recommendation clearly shows that adding new auxiliary data even with the same learning function can enhance the overall recommendation performance. 

Even though NHR-text system improves the recommendation quality, it underperforms NHR-categorical because of its model complexity. Besides the inevitable large size of the embedding layer, the Kariyer dataset is extremely sparse and interaction data is not enough to feed such a network in fact. With more data, we expect to have more contribution from text data. 

The last but not the least, the results are more promising for the job recommendation. Since Kariyer dataset suffers from a severe sparsity and a high frequency of cold-starts, the auxiliary data and the cooperation of models can fill in this information shortage about user preferences. 

\section{Conclusion}
\label{sec:conclusion}
In this work, we explored DNNs for hybrid recommender systems 
. We devised a general framework NHR that model user-item interactions by combining auxiliary and historical data. We showed that every variation of NHR outperforms state-of-the-art collaborative filtering methods as expected, but NHR also gives us the chance to alleviate deficiencies to be dependent on single aspects or data sources. It does not require to train complete architecture from scratch. Instead, it allows self-sufficient recommender models to speak for themselves by a weighting process which learns the capabilities of its components.

In the next phase of the study, we would like to test our approach on explicit datasets and use pre-trained vector space models such as document vectors for text features since learning of embedding layers directly effects the model complexity and training time. Since average pooling leads to the loss of sequential property of natural language texts, we would like to improve our text models by using more elaborated architectures such as LSTMs and CNNs to exploit sequence information and interrelation of words.
\subsection*{Acknowledgements}
This study is part of the research project (Project No:5170032) supported by the Scientific and Technological Research Council of Turkey (T\"UB\.{I}TAK). The authors would like to thank Istanbul Technical University for their financial support under the project BAP-40737.

%
%
%
\bibliographystyle{splncs04}
\bibliography{bibliography}
\end{document}